\documentclass[twocolumn,showpacs,prl,10pt,aps,floatfix]{revtex4-1}
\usepackage[english]{babel}
\usepackage{graphicx}
\usepackage[colorlinks,linkcolor=blue,citecolor=blue,urlcolor=blue]{hyperref}
\usepackage[table]{xcolor}
\usepackage{amsmath}
\usepackage{soul}
\usepackage{setspace}

\begin{document}

\title{Spin dynamics of the block orbital-selective Mott phase} 

\author{J. Herbrych$^{1,2}$}
\email{jherbryc@utk.edu}
\author{N. Kaushal$^{1,2}$}
\author{A. Nocera$^{1,3}$}
\author{G. Alvarez$^{2,3}$}
\author{A. Moreo$^{1,2}$}
\author{E. Dagotto$^{1,2}$}
\affiliation{$^{1}$ Department of Physics and Astronomy, University of Tennessee, 
Knoxville, Tennessee 37996, USA}
\affiliation{$^{2}$ Materials Science and Technology Division, Oak Ridge National 
Laboratory, Oak Ridge, Tennessee 37831, USA}
\affiliation{$^{3}$ Computational Sciences and Engineering Division and Center for 
Nanophase Materials Sciences, Oak Ridge National Laboratory, Oak Ridge, Tennessee 
37831, USA}

\date{\today}

\begin{abstract}
Iron-based superconductors display a variety of magnetic phases originating in the competition between electronic, orbital, and spin degrees of freedom. Previous theoretical investigations of the multi-orbital Hubbard model in one dimension revealed the existence of an orbital-selective Mott phase (OSMP) with block spin order. Recent inelastic neutron scattering (INS) experiments on the BaFe$_2$Se$_3$ ladder compound confirmed the relevance of the block-OSMP. Moreover, the powder INS spectrum reveled an unexpected structure, containing both low-energy acoustic and high-energy optical modes. Here we present the theoretical prediction for the dynamical spin structure factor within a block-OSMP regime using the density-matrix renormalization group method. In agreement with experiments we find two dominant features: low-energy dispersive and high-energy dispersionless modes. We argue that the former represents the spin-wave-like dynamics of the block ferromagnetic islands, while the latter is attributed to a novel type of local on-site spin excitations controlled by the Hund coupling.
\end{abstract}

\maketitle


Inelastic neutron scattering (INS) measurements are crucial for the study of quantum 
magnetism in condensed matter physics. This powerful experimental technique provides 
detailed information of momentum and energy resolved spin excitations. The importance 
of INS studies is best illustrated in the case of high critical temperature 
superconductors. Shortly after the discovery of the copper-oxide compounds it became 
evident that the standard BCS theory of the electron-phonon coupling could not explain 
the experimental findings. Simultaneously, INS results showed that superconductivity 
appears in close proximity to the antiferromagnetic (AFM) ordering of 
$S=1/2$ Cu$^{2+}$ moments providing robust evidence that the new pairing mechanism is 
based on spin fluctuations~\cite{cuprates-INS}.

The discovery of iron-based superconductors (FeSC) added an extra complication to this 
``simple'' picture. Although the phase diagrams of Cu-based and Fe-based materials are 
qualitatively similar~\cite{Basov2011}, there are important conceptual differences. 
The most significant are in the minimal models that describe the materials 
\cite{Daghofer2010,Fernandes2017}. While cuprates have a single Fermi surface (FS), the 
iron-based compounds have a complicated FS with electron and hole pockets originating 
in the five $3d$ orbitals of iron. As a consequence, the FeSC have to be described by 
means of multi-orbital Hubbard models, involving not only a standard Hubbard $U$ 
repulsion but also a Hund coupling. The competition between electronic, orbital, and 
spin degrees of freedom can lead to many exotic magnetic phases 
\cite{Yin2010,Lumsden2010,Dai2012,dagotto2013,Li2014,Bascones2016}. 

Past experience in cuprates showed that the analysis of lower dimensional systems, 
such as chains and ladders, can provide useful information to better contrast theory 
with experiments~\cite{dagotto1996}. One reason is that theoretical many-body 
calculations based on model Hamiltonians can be accurately performed in one dimension, 
particularly numerically. For this reason, it was exciting when a one-dimensional 
family of compounds containing two-leg ladders was unveiled also in the 
iron-superconductors context. Specifically, we refer to the low-dimensional FeSC in 
the 123 family, $A$Fe$_2$$X_3$, where $A$ are alkali metals $A$=K, Ba, Rb, Cs, and $X$ 
are chalcogenides $X$=S, Se. These compounds are build of double chains (i.e. they are 
ladders) of edge sharing Fe$X_4$ tetrahedra~\cite{Takubo2017}. Recently, a 
superconducting state was identified under pressure for BaFe$_2$S$_3$ 
\cite{Takahashi2015,Yamauchi2015} and BaFe$_2$Se$_3$~\cite{BNL-SC,Zhang2018}. The 
pressure-dependent phase diagram of these materials resembles that of copper-oxide 
ladders, e.g., the telephone number compound 
Sr$_{14-x}$Ca$_x$Cu$_{24}$O$_{41}$~\cite{Uehara1996}. Similar to their copper oxide 
counterparts, the iron-123 family is insulating at ambient pressure. This behavior is 
unusual since, unlike the cuprates, the parent compounds of FeSC are typically bad 
metals. In addition, it was argued that orbital-selective Mott physics 
(OSMP)~\cite{georges2013} is consistent with results for BaFe$_2$Se$_3$ 
\cite{Caron2012,Rincon2014-1,Dong2014,Mourigal2015}.  Within such a phase, itinerant 
and localized conduction electrons coexist.

\begin{figure*}[!ht]
\includegraphics[width=1.00\textwidth]{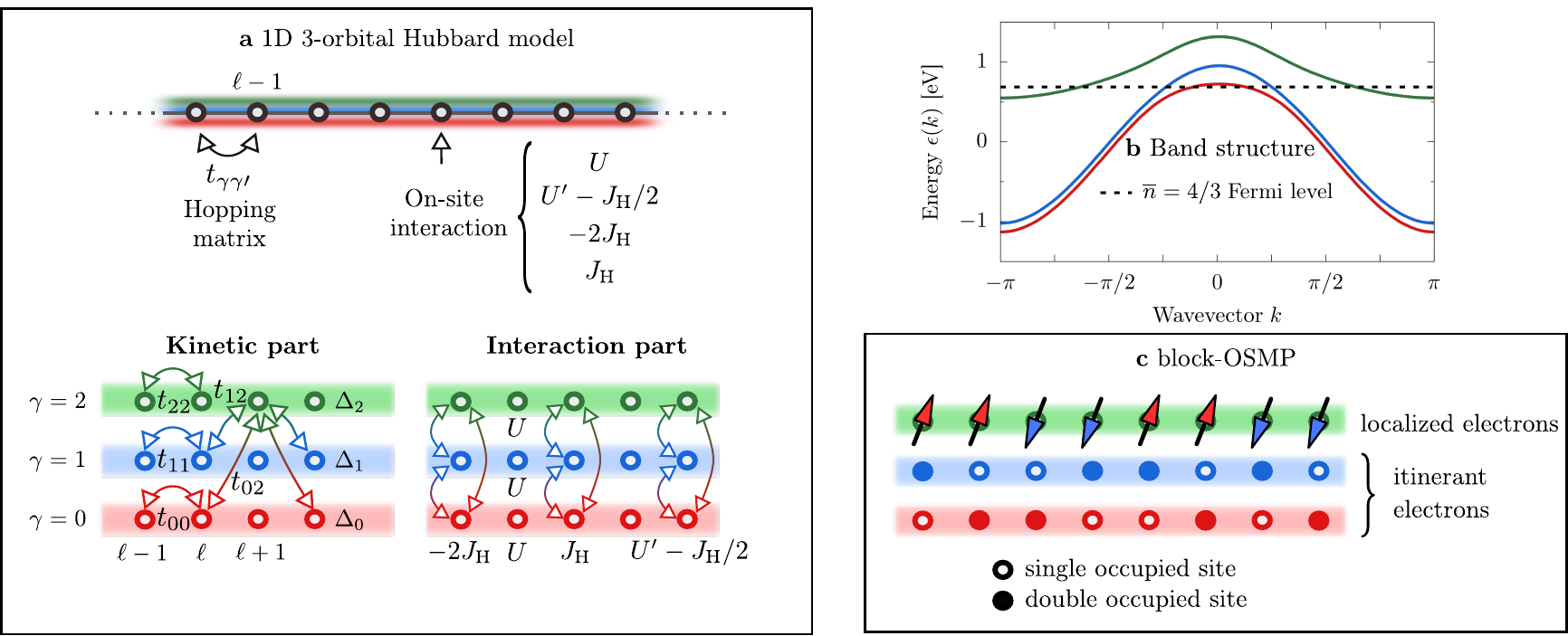}
\caption{{\bf Schematic representation of the Hamiltonian.} 
(a) Three-orbital Hubbard model on a one-dimensional lattice geometry (see text for 
details). (b) Band structure of Hamiltonian Eq.~(\ref{hamkin}). Note that due 
to the hybridization $t_{\gamma\gamma^\prime}\ne0$ for ${\gamma \neq \gamma^\prime}$, 
the band numbers do not correspond 
directly to the orbital numbers. (c) Schematic representation of the block orbital 
selective Mott phase. The pattern of single and double occupied sites in the itinerant 
electrons is meant to be random, representing pictorially the not-localized nature of 
those orbitals.}
\label{fig:sche}
\end{figure*}

It should be remarked that INS experiments on 123 materials have been performed up to 
now only on powder samples and, as a consequence, detailed data of the momentum 
dependence of the spin excitations over the whole Brillouin zone is not yet available. 
Nevertheless, the static $(\pi,0)$ stripe AFM order -- with ferromagnetic rungs and 
antiferromagnetic legs -- was identified for BaFe$_2$S$_3$~\cite{wang2017}, 
RbFe$_2$Se$_3$~\cite{wang2016}, CsFe$_2$Se$_3$~\cite{Hawai2015,Chi2016}, and also for 
KFe$_2$S$_3$~\cite{Caron2012}. However, in the special case of BaFe$_2$Se$_3$ 
remarkably an exotic block magnetism was found 
\cite{Caron2011,Caron2012,Nambu2012,Mourigal2015,Zhang2018} involving 
antiferromagnetically coupled ferromagnetic islands made of 2$\times$2 iron clusters. 
This unusual magnetic state was also observed in the vicinity of superconductivity 
\cite{Guo2010,Shermadini2011,Ye2011} in two-dimensional (2D) materials with 
$\sqrt{5}\times\sqrt{5}$ ordered iron vacancies, such as 
Rb$_{0.89}$Fe$_{1.58}$Se$_{2}$~\cite{wang2011} and 
K$_{0.8}$Fe$_{1.6}$Se$_{2}$~\cite{Bao2011,You2011,Yu2011}. In addition, for 
BaFe$_2$Se$_3$~\cite{Mourigal2015}, BaFe$_2$S$_3$~\cite{wang2017}, and 
RbFe$_2$Se$_3$~\cite{wang2016} the INS revealed the existence of low-energy acoustic 
and high-energy optical modes separated by an energy gap. 
It is important to 
remark that the generic features of the INS spectra of the aforementioned compounds are 
similar, but the physical origin of the acoustic modes can differ significantly - 
these  modes reflect on the long-distance properties of the magnetic order in the 
system. Moreover, the origin and characteristics of the optical modes, that 
are induced by short-distance properties, have not been clarified so far.

In this work, we will address the spin dynamical properties of the exotic block 
magnetic state found in BaFe$_2$Se$_3$. The static (time independent) properties of 
this phase were previously qualitatively studied in Ref.~\cite{Rincon2014-1} via a 
three-orbital Hubbard model in one-dimension (1D) that unveiled an OSMP regime. Here, 
we will use the same Hamiltonian to investigate the momentum and energy resolved spin 
dynamics. To test the general features of our findings we present also results 
obtained in a quasi-1D ladder geometry. In agreement with experimental findings we have observed 
two distinct modes of spin excitations: a low-energy dispersive mode and high-energy 
dispersionless optical modes. The low-energy acoustic mode reveals the frustrated 
nature of the block magnetism which can be described by a spin $J_1$-$J_2$ 
Hamiltonian. On the other hand, we argue that the optical mode is controlled by local orbital 
physics and it cannot be properly captured by a Heisenberg-like model. The main 
features of our analysis are simple and generic and should characterize any 
multi-orbital model as long as its ground state is in a magnetic block phase.

\section{Results}

{\bf Model and observables.} 
We will focus on a specific three-orbital Hubbard model on a one-dimensional lattice, 
but our conclusions are generic for a broad group of models and materials in the OSMP 
magnetic block-phase regime. As mentioned before, the model chosen was previously 
studied with regards to its time-independent properties, and it is known that it 
displays an OSMP regime in the ground state~\cite{Rincon2014-1}. The kinetic part of 
the Hamiltonian, $H_{\mathrm{kin}}$, is defined as:
\begin{equation}
H_{\mathrm{kin}}=-\sum_{\ell,\sigma,\gamma,\gamma^\prime}
t_{\gamma\gamma^\prime}\left(
c^{\dagger}_{\ell,\gamma,\sigma}
c^{\phantom{\dagger}}_{\ell+1,\gamma^\prime,\sigma}+\mathrm{H.c.}\right)
+\sum_{\ell,\gamma,\sigma}\Delta_{\gamma} n_{\ell,\gamma,\sigma}\,,
\label{hamkin}
\end{equation}
where $c^{\dagger}_{\ell,\gamma,\sigma}$ creates an electron with spin 
$\sigma=\{\uparrow,\downarrow\}$ at orbital $\gamma=\{0,1,2\}$ and site 
$\ell=\{1,\dots,L\}$ of a 1D chain. 
$n_{\ell,\gamma,\sigma}=c^{\dagger}_{\ell,\gamma,\sigma} c^{\phantom{\dagger}}_{\ell,\sigma,\gamma}$ 
is the local $(\ell,\gamma)$ electron density with spin $\sigma$. Note that another 
common labeling of these orbitals could be based on the canonical $t_{2g}$ manifold, 
i.e. $\{yz\,,xz\,,xy\}$, respectively. $t_{\gamma\gamma^\prime}$ denotes a symmetric 
hopping amplitude matrix defined in the orbital space $\gamma$: $t_{00}=t_{11}=-0.5$, 
$t_{22}=-0.15$ , $t_{02}=t_{12}=0.1$ and $t_{01}=0$, all in $\mathrm{eV}$ units 
[Fig.~\ref{fig:sche}(a) displays a schematic representation of the Hamiltonian]. The 
crystal-field splitting is set to $\Delta_{0}=-0.1$, $\Delta_{1}=0$, and 
$\Delta_{2}=0.8$, also in $\mathrm{eV}$ units. The total kinetic-energy bandwidth is 
$W=2.45\,\mathrm{eV}$. These phenomenological values of parameters were chosen 
before~\cite{Rincon2014-1} to reproduce qualitatively the band structure properties of 
higher dimensional selenides at an electronic density $\overline{n}=4/3$ per orbital, 
namely an electron-like pocket at $k=0$ and hole-like pockets at $k=\pm\pi$
(see Fig.~\ref{fig:sche}(b), and also Ref.~\cite{luo2013} and references 
therein). It should be pointed out that the existence of an OSMP highlights the 
striking orbital sensitivity on electron correlations in multi-orbital Hubbard models, 
and its presence is not limited to our use of 1D geometries nor to our choice of 
$t_{\gamma\gamma^\prime}$ hoppings. For example, the OSMP was proven to be 
relevant~\cite{Yu2013} for 2D alkaline iron selenides as well, with and without 
$\sqrt{5}\times\sqrt{5}$ ordered vacancies. We wish to emphasize that our predictions 
primarily depend on the existence of an OSMP magnetic block-phase state, rather than on 
the  details of the Hamiltonian that leads to its stabilization. In this context, we 
believe that our results are universal for iron-based superconductors. To support this 
claim, we will present calculations for several models showing that all the many 
reported results lead essentially to the same qualitative conclusions.

The interaction portion of the Hamiltonian $H_{\mathrm{int}}$ is given by
\begin{eqnarray}
&&H_{\mathrm{int}}=U\sum_{\ell,\gamma}
n_{\ell,\gamma,\uparrow}n_{\ell,\gamma,\downarrow}
+(U^\prime-J_{\mathrm{H}}/2)\sum_{\ell,\gamma<\gamma^\prime}
n_{\ell,\gamma}n_{\ell,\gamma^\prime}\nonumber\\
&-&2J_{\mathrm{H}}\sum_{\ell,\gamma<\gamma^\prime}S_{\ell,\gamma}S_{\ell,\gamma^\prime}
+J_{\mathrm{H}}\sum_{\ell,\gamma<\gamma^\prime}
\left(P^{+}_{\ell,\gamma}P^{\phantom{+}}_{\ell,\gamma^\prime}
+\mathrm{H.c.}\right)\,,
\label{hamint}
\end{eqnarray}
where $n_{\ell,\gamma}=\sum_{\sigma}n_{\ell,\gamma,\sigma}$, the local spin 
$(\ell,\gamma)$ is $S_{\ell,\gamma}=(1/2)\sum_{a,b}c^{\dagger}_{\ell,\gamma,a}\sigma^{ab}c^{\phantom{\dagger}}_{\ell,\gamma,b}$ 
(with $\sigma^{ab}$ as a Pauli spin matrices), and 
$P^{\phantom{+}}_{\ell,\gamma}=c_{\ell,\uparrow,\gamma}c_{\ell,\downarrow,\gamma}$ is 
the pair-hopping. We will consider an SU(2) symmetric system, i.e., 
$U^{\prime}=U-2J_{\mathrm{H}}$, where $U$ stands for the on-site same-orbital repulsive Hubbard 
interaction. Finally, we set the Hund coupling to $J_{\mathrm{H}}=U/4$, a value widely used 
before and considered to be realistic for Fe-based materials~\cite{Haule2009,Luo2010}. 
We refer the interested reader to Refs.~\cite{Rincon2014-1,Rincon2014-2,Kaushal2017} 
for details of the $J_{\mathrm{H}}$-$U$ phase diagram of the above Hamiltonian. Here, if not 
stated differently, we will use $U/W=0.8$ where previous studies 
found~\cite{Rincon2014-1} a block-OSMP , i.e. antiferromagnetically (AFM) coupled 
ferromagnetic (FM) blocks (magnetic unit cells), 
$\uparrow\uparrow\downarrow\downarrow\uparrow\uparrow\downarrow\downarrow$, in the 
localized orbital $\gamma=2$ [see Fig.~\ref{fig:sche}(c)]. Note that the 
block order is usually studied in the context of Heisenberg-like spin Hamiltonians 
(such as dimerized~\cite{Mourigal2015,wang2011} or $J_1$--$J_2$ models~\cite{ren2012}). 
Here, the block phase is a consequence of nontrivial electronic correlations within the
OSMP phase. Since the latter is a feature of multi-orbital systems that cannot be analyzed
using purely spin systems, we believe that 
our setup is more suitable for the study of iron-based materials.

In this work, we will investigate the zero-temperature frequency $\omega$-dependent 
spin structure factor (SSF) $S(q,\omega)$, defined as the Fourier transform of the 
real-space total (on-site, $S_{\ell}=\sum_{\gamma}S_{\ell,\gamma}$) spin correlation 
functions (see Methods). Furthermore, we will study the contributions from the 
individual orbitals to the total SSF, i.e. $S_{\gamma\gamma^\prime}(q,\omega)$. 
$\gamma=\gamma^\prime$ denotes the spin fluctuations within each of the orbitals, 
while $\gamma\ne\gamma^\prime$ are spin fluctuations between different orbitals. As a 
consequence 
$S(q,\omega)=\sum_{\gamma}S_{\gamma\gamma}(q,\omega)+\sum_{\gamma\ne\gamma^\prime}S_{\gamma\gamma^\prime}(q,\omega)$. 
From the experimental perspective, only the total SSF has a 
meaning~\cite{Nicholson2011} because neutrons couple to electrons in all orbitals in 
neutron scattering experiments. However, the theoretical investigations of 
orbital-resolved SSF can provide further insight into the OSMP physics.

The Hamiltonians are diagonalized via the DMRG method, where the dynamical correlation 
functions are obtained with the help of dynamical DMRG techniques (see Methods and 
Supplementary Note~1 for details of the numerical simulations).

\begin{figure}[!ht]
\includegraphics[width=1.0\columnwidth]{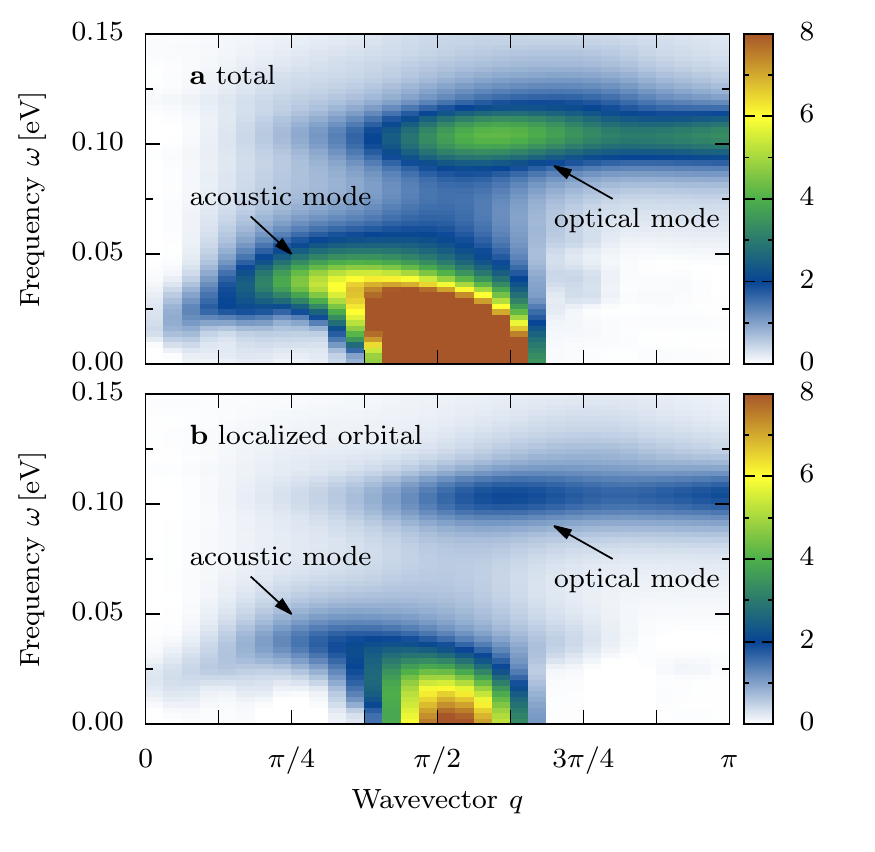}
\caption{{\bf Dynamical spin structure factor (SSF).} (a) 
Total SSF, $S(q,\omega)$, and (b) SSF of the localized orbital, $S_{22}(q,\omega)$. 
Both results exhibit a low-energy acoustic and a high-energy optical modes. Note that 
the spectral weight of the localized orbital, $S_{22}$, constitutes $\sim50\%$ of the 
total SSF weight. The results were obtained using a dynamical DMRG method with 
parameters $L=16$ ($48$ orbitals), $M=800$, $\delta\omega=0.005\,\mathrm{eV}$, and 
$\eta/\delta\omega=2$.}
\label{fig:U08}
\end{figure}

{\bf Dynamical spin structure factor.} 
In Fig.~\ref{fig:U08} we present one of the main results of our effort: the 
frequency-momentum dependence of the dynamical SSF in the block-OSMP phase (i.e. at 
$U/W=0.8$). Panel (a) depicts the total SSF, $S(q,\omega)$, while panel (b) shows only 
the contribution from the localized orbital, $S_{22}(q,\omega)$. Several conclusions 
can be obtained directly from the presented results: (i) A robust contribution to the 
total SSF arises from the localized orbital. Moreover, all the qualitative features of 
$S(q,\omega)$ are already present in $S_{22}(q,\omega)$. In fact, $S(q,\omega)$ and 
$S_{22}(q,\omega)$ become almost indistinguishable
if normalized by the local magnetic moment squared (i.e. $S^2=3/4$ for the $S=1/2$ 
localized electron, and $S^2=2$ for the total moment~\cite{Rincon2014-1}). (ii) The 
energy range for the spin dynamics is much smaller when compared with the energy 
bandwidth $W=2.45\,\mathrm{eV}$ of the Hamiltonian. (iii) Clearly the dynamical SSF 
has two distinct modes: a low-frequency, $\omega\lesssim \omega_c=0.08\,\mathrm{eV}$, 
dispersive (acoustic) band and a high-frequency, $\omega\sim0.11\,\mathrm{eV}$, 
dispersionless (optical) band. Similar results were previously reported 
experimentally in INS investigations of BaFe$_2$Se$_3$~\cite{Mourigal2015} (with 
$2\times2$ FM blocks), BaFe$_2$S$_3$~\cite{wang2017} and 
RbFe$_2$Se$_3$~\cite{wang2016} (with $2\times1$ FM blocks). The different types of 
blocks in the INS investigations, and the similarity of results between neutrons and 
our calculations, suggest that our results apply to a broad variety of iron 
chalcogenides. Moreover, the INS measurements where performed on powder samples and, 
as a consequence, no detailed analysis of the spin excitations over all crystal 
momenta $q$ (over the whole Brillouin zone) have been reported. In this respect, our 
results define clear theoretical predictions on what future single-crystal experiments 
should display.

\begin{figure}[!ht]
\includegraphics[width=1.0\columnwidth]{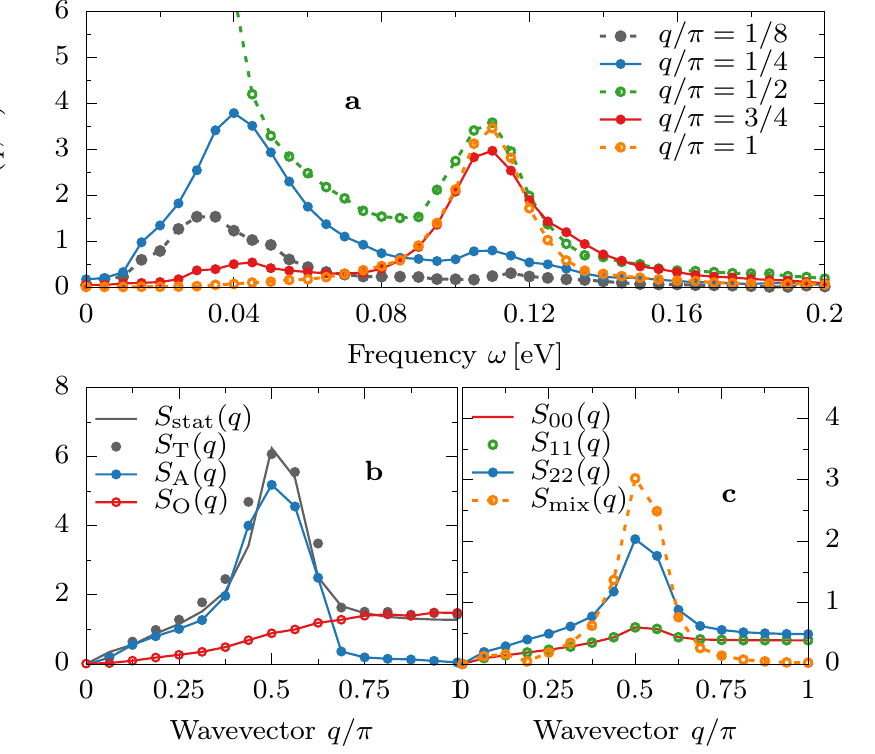}
\caption{{\bf Frequency and momentum dependence.} (a) Finite momentum cuts, 
$q/\pi=1/8,1/4,1/2,3/4,1$, of the dynamical SSF. (b) Total static SSF obtained as an 
expectation value in the $|\mathrm{GS}\rangle$, $S_{\mathrm{stat}}(q)$ (black line), 
and also via the $\omega$ integration of the dynamical SSF, $S_{\mathrm{T}}(q)$ (black 
points). In the same panel we present also the contributions to the static SSF from 
acoustic and optical modes, $S_{\mathrm{A}}(q)$ and $S_{\mathrm{O}}(q)$, respectively. 
(c) Contribution to the static SSF $S_{\gamma\gamma^\prime}(q)$: 
$\gamma=\gamma^\prime$ represents the SSF component for each of the orbitals, while 
$S_{\mathrm{mix}}(q)=\sum_{\gamma\ne\gamma^\prime}S_{\gamma\gamma^\prime}(q)$ 
represents the sum of the inter-orbital contributions. The system parameters are the 
same as in Fig.~\ref{fig:U08}.}
\label{fig:omega}
\end{figure}

\begin{figure*}[!ht]
\includegraphics[width=0.8\textwidth]{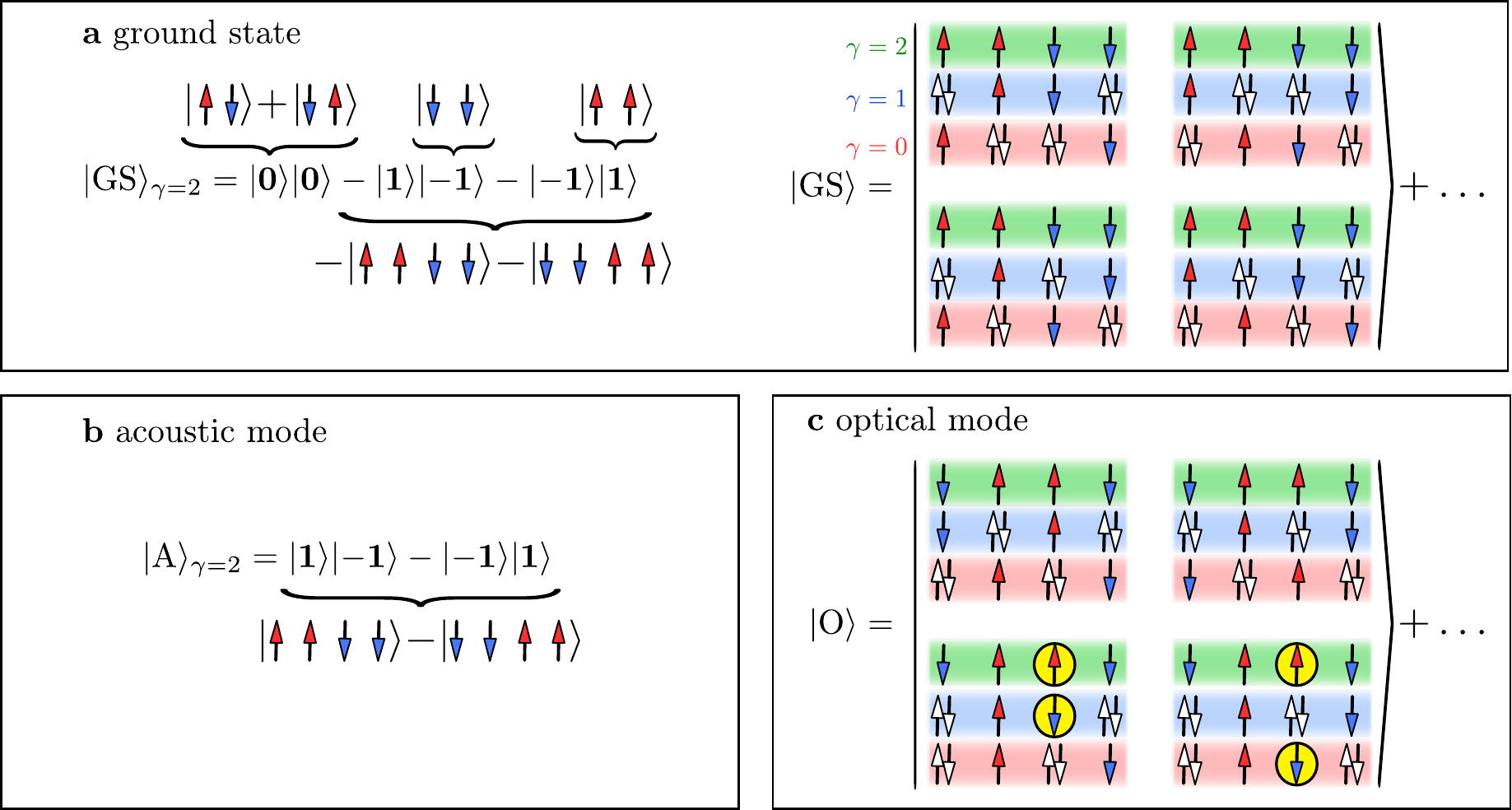}
\caption{{\bf States relevant for the dynamical SSF.} 
Spin configuration in the localized orbital ($\gamma=2$) (see text for details) of the 
(a) $|\mathrm{GS}\rangle$ (singlet) and (b) $|\mathrm{A}\rangle$ state (triplet) 
contributing to the acoustic mode. (c) Schematic representation of particle 
configuration of all orbitals of the $|\mathrm{GS}\rangle$ and optical triplet 
$|\mathrm{O}\rangle$. Circles represent pairs of antiferromagnetically aligned spins 
which break the Hund's rule.}
\label{fig:conf}
\end{figure*}

In Fig.~\ref{fig:omega}(a) we present the $\omega$ dependence of the total SSF at 
special values of the momenta $q$. It is evident that the acoustic mode is strongly 
momentum dependent in the range $0<q/\pi\lesssim 1/2$, while it reduces drastically 
its intensity for $q/\pi>1/2$. To understand these results, we can reanalyze the SSF 
spectrum using two-sites as a rigid block, namely creating an effective magnetic 
unit cell of FM-blocks with momentum $\widetilde{q}$. The acoustic mode as a function 
of $\widetilde{q}$ then is located between $0<\widetilde{q}<\pi$, resembling a gapless 
continuum of spin excitations. Such an interpretation is consistent with 
``collective'' spin waves based on FM blocks. On the other hand, the high-energy 
optical contribution is $q$-independent for $q/\pi\gtrsim1/2$, with vanishing 
intensity in the $q\to 0$ limit. As discussed later, this mode can be associated to 
local (on-site) spin excitations affecting the Coulomb potential portion of the 
Hamiltonian, independently of the dimensionality of the system. The $q$-dependence of 
both modes is also clearly visible in the static SSF obtained from the energy 
integration of the dynamical SSF, i.e. 
$S_{\alpha}(q)=(1/\pi)\int\mathrm{d}\omega\,S(q,\omega)$. In Fig.~\ref{fig:omega}(b) 
we present the acoustic ($\alpha=\mathrm{A}$) and optical ($\alpha=\mathrm{O}$) 
contribution to the total ($\alpha=T$) static SSF, coming from the integration over 
the frequency ranges $0<\omega<\omega_{c}$, $\omega_{c}<\omega<\infty$, and 
$0<\omega<\infty$, respectively. From the dynamical SSF spectra, it is evident that 
$S_{\mathrm{O}}(q)$ provides the sole contribution to the total static SSF for 
momentum $0.75<q/\pi<1$. As a consequence, at least within a block-OSMP state it is 
remarkable that already in the static SSF one can observe the clear presence of an 
optical mode, a novel result which is intrinsic of block phases to our knowledge. 
In the same panel, we 
also present the total static SSF independently obtained from the expectation value of 
the ground state (GS), i.e., 
$S_{\mathrm{stat}}(q)=\langle \mathrm{GS}|S_{q} \cdot S_{-q}|\mathrm{GS}\rangle$, 
where $S_q$ is the Fourier transform of the $S_\ell$ operators for the same system 
size $L$. The agreement between $S_{\mathrm{stat}}(q)$ and $S_{\mathrm{T}}(q)$ serves 
as nontrivial accuracy test of the dynamical DMRG method, since the former can be 
obtained with much higher accuracy.

{\bf Orbital contribution.} 
Before addressing the optical and acoustic modes in more detail, we will comment on 
the orbital $\gamma$ contribution to $S(q,\omega)$. As already shown in 
Fig.~\ref{fig:U08}, the main contribution to the total SSF originates in the localized 
orbital $\gamma=2$. Our results [see Fig.~\ref{fig:omega}(c)] indicate that the spin 
fluctuations for the itinerant electrons (orbitals $\gamma=0$ and $\gamma=1$) follow 
the behavior of the localized orbital. As argued below, this is a consequence of the 
Hund coupling which aligns ferromagnetically spins at different orbitals. However, the 
nature of these orbitals is metallic and magnetic moments are not well formed. As a 
consequence, the spectral weight of the total itinerant contribution (2 orbitals) is 
approximately the same as the localized (1 orbital). On the other hand the 
inter-orbital SSF $S_{\gamma\ne\gamma^\prime}$ have a large contribution only to the 
acoustic mode, especially near the $q/\pi=1/2$ point. 

\begin{figure}[!ht]
\includegraphics[width=1.0\columnwidth]{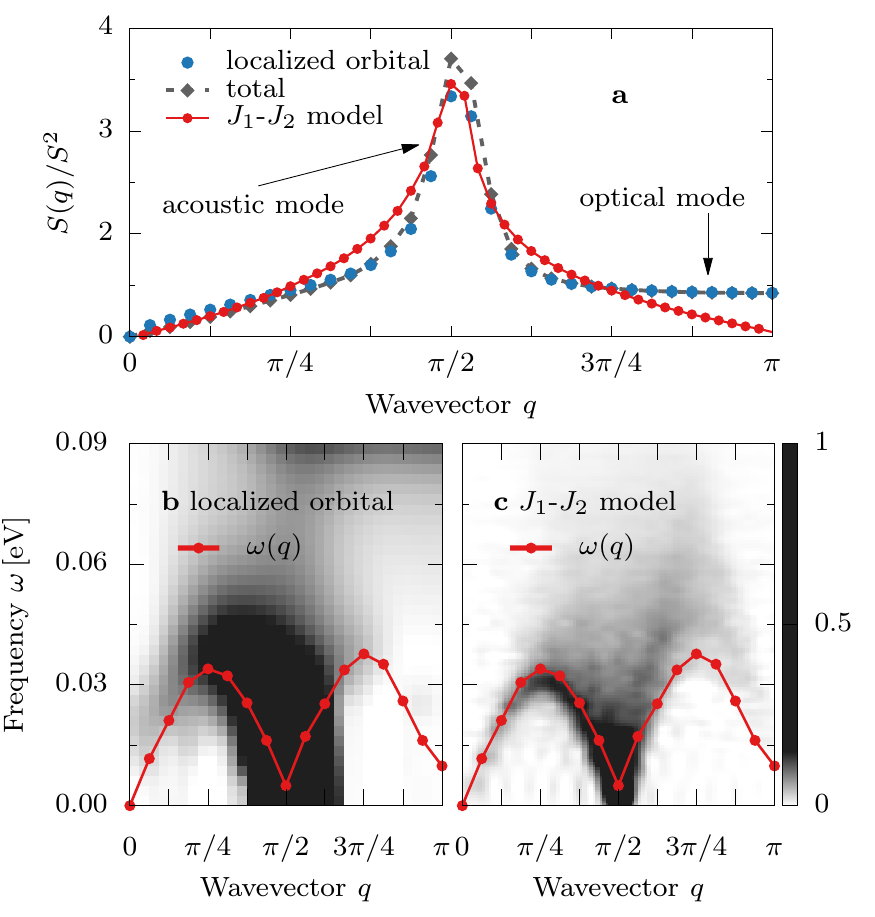}
\caption{{\bf Comparison with the $J_1$--$J_2$ model.} (a) Comparison of the static 
SSF $S(q)$ corresponding to the multi-orbital system vs results for the $J_1$--$J_2$ 
model with $J_2/|J_1|=1$, calculated for $L=32$ and $L=48$, respectively. $S(q)$ is 
normalized by the local magnetic moment squared $S^2=S(S+1)$, where $S^2=3/4$ for the 
localized orbital and the $J_1$--$J_2$ result, and $S^2=2$~\cite{Rincon2014-1} for the 
total SSF. (b) Acoustic mode of the dynamical SSF within the block-OSMP phase [the 
same results as in Fig.~\ref{fig:U08}(b)] compared against the dispersion relation 
$\omega(q)$ of the $J_1$--$J_2$ model with $|J_1|=J_2=0.6J_{\mathrm{eff}}$, where the 
effective spin exchange energy scale is set to $J_{\mathrm{eff}}=4\,t_{22}^2/U$. The 
latter is obtained with the help of Lanczos diagonalization on a chain of $32$ sites 
with periodic boundary conditions. (c) Dynamical SSF of the $J_1$--$J_2$ model with 
$|J_1|=J_2=0.6J_{\mathrm{eff}}$ as calculated using DMRG for a $L=48$ chain.}
\label{fig:disp}
\end{figure}

{\bf Acoustic mode.}
Consider now the properties of the acoustic mode. Motivated by the results presented 
above, with the main contribution to the SSF arising from the localized orbital, we 
express the eigenstates in terms of the basis states of localized orbital 
$|\,\cdot\,\rangle_{\gamma=2}$ (see Methods section). Since the electrons are indeed 
localized with occupation $n_{\gamma=2}=1$~\cite{Rincon2014-1} in the OSMP, in the 
low-energy portion of the spectrum the basis states with empty and double occupied 
orbital $\gamma=2$ should not be present. Within such a representation the GS of the 
block-OSMP phase can be identified as a superposition of 
$|\uparrow\uparrow\downarrow\downarrow\rangle_{\gamma=2}$ and 
$|\downarrow\downarrow\uparrow\uparrow\rangle_{\gamma=2}$ states which constitutes 
$\sim50\%$ of the true GS. One can improve further the qualitative description by 
investigating a simple toy model. Let us consider two FM coupled $S=1/2$ spins as one 
$S=1$ object, i.e. 
$|\mathbf{1}\rangle=|\uparrow\uparrow\rangle_{\gamma=2}$, $|\mathbf{-1}\rangle=|\downarrow\downarrow\rangle_{\gamma=2}$, and \mbox{$|\mathbf{0}\rangle=1/\sqrt{2}(|\uparrow\downarrow\rangle_{\gamma=2}+|\uparrow\downarrow\rangle_{\gamma=2})$}. In this 
setup, a $4$-site $S=1/2$ system reduces to two antiferromagnetically coupled $S=1$ 
spins. The ground state of the latter is simply
\begin{equation}
|\mathrm{GS}\rangle_{\gamma=2} =c_a|\mathbf{0}\rangle|\mathbf{0}\rangle-
c_b\Big(|\mathbf{1}\rangle|\mathbf{-1}\rangle+
|\mathbf{-1}\rangle|\mathbf{1}\rangle\Big)\,,
\label{toygs}
\end{equation}
where $c_a=c_b=1/\sqrt{3}$ [see Fig.~\ref{fig:conf}(a) for a schematic 
representation]. Note that the above state, in agreement with numerics, is a singlet. 
The last two terms of Eq.~\ref{toygs} correspond to the ``perfect'' block order, i.e., 
\mbox{$|\uparrow\uparrow\downarrow\downarrow\rangle_{\gamma=2}+|\downarrow\downarrow\uparrow\uparrow\rangle_{\gamma=2}$}, 
while the first term depicts the $x$--$y$ component of the block order,
\begin{eqnarray}
|\mathbf{0}\rangle|\mathbf{0}\rangle=\frac{1}{2}
\Big(|\uparrow\downarrow\uparrow\downarrow\rangle_{\gamma=2}+
|\downarrow\uparrow\downarrow\uparrow\rangle_{\gamma=2}\Big.\nonumber\\
\Big.+|\uparrow\downarrow\downarrow\uparrow\rangle_{\gamma=2}+
|\downarrow\uparrow\uparrow\downarrow\rangle_{\gamma=2}\Big)\,.
\end{eqnarray}
Our $L=4$ Lanczos investigation of the full Hamiltonian (\ref{hamkin}-\ref{hamint}) 
indicates that such a state has coefficients equal to $\widetilde{c}^2_a\simeq1/6$ and 
$\widetilde{c}_b^2\simeq1/4$, which yields now a better overlap, $\sim70\%$, with the 
true GS. Finally, the first excited state - contributing to the acoustic mode - can be 
identified as a triplet of the form 
\mbox{$|\mathrm{A}\rangle_{\gamma=2}=\widetilde{c}_{A} (|\uparrow\uparrow\downarrow\downarrow\rangle_{\gamma=2}-|\downarrow\downarrow\uparrow\uparrow\rangle_{\gamma=2})$} 
where $\widetilde{c}_{A}^2\simeq4/9$ [see Fig.~\ref{fig:conf}(b)]. This large overlap 
of \mbox{$|\mathrm{A}\rangle_{\gamma=2}$} with the full solution is also captured by 
the toy model since 
$|\mathbf{1}\rangle|\mathbf{-1}\rangle-|\mathbf{-1}\rangle|\mathbf{1}\rangle$ is one 
of the first excitations in our two-site $S=1$ problem. Note that the above 
description of the $|\mathrm{GS}\rangle_{\gamma=2}$ ($|\mathrm{A}\rangle_{\gamma=2}$) 
as a spin singlet (triplet) is not obvious from the signs of the localized orbital 
basis representation. While the above states capture the essence of the problem, the 
itinerant orbitals have to be included in the description to account for the true 
nature of the singlet-triplet excitation.

Although simplified, descriptions such as those above of the low-energy spectrum can 
yield nontrivial consequences.  A similar ground state to our 
$|\mathrm{GS}\rangle_{\gamma=2}$ with $\pi/2$ pitch angle was previously 
observed in the frustrated ferromagnetic $S=1/2$ $J_1$--$J_2$ Heisenberg model with 
ferromagnetic $J_1$ and antiferromagnetic 
$J_2$~\cite{Bursill95,enderle2010,ren2012,onishi2015,onishi2015-2}. In 
Fig.~\ref{fig:disp}(a) we present a comparison of the multi-orbital system 
Eqs.~(\ref{hamkin}-\ref{hamint}) SSF vs $J_1$-$J_2$ results obtained for 
$J_2/|J_1|=1$. Within the latter the dynamical SSF yields a continuum of excitations 
with maximum intensity at $q/\pi=1/2$ and vanishing intensity in the $q/\pi\to1$ 
limit. In fact, the dynamical SSF of the $J_1$--$J_2$ model is very similar to the 
acoustic mode found in our multi-orbital system, i.e. compare panels (b) and (c) of 
Fig.~\ref{fig:disp}. To strengthen this argument, in Fig.~\ref{fig:disp}(b,c) we 
present the dynamical SSF factor plotted against the quantum dispersion relation of 
the $J_1$--$J_2$ model $\omega(q)=\epsilon_q-\epsilon_{\mathrm{GS}}$ where 
$\epsilon_q$ is the energy of the lowest eigenstate at a given $q$. To match the 
energy scales we set $|J_1|=J_2=0.6J_{\mathrm{eff}}$ where 
$J_{\mathrm{eff}}=4\,t_{22}^2/U$ is the natural superexchange scale within the 
localized orbital, as a crude approximation. As clearly shown in 
Fig.~\ref{fig:disp}(b), $\omega(q)$ quantitatively captures the main features of the 
acoustic portion of the spectrum. 

We remark that the present comparison with 
the $J_1$--$J_2$ model is at a phenomenological level, since this effective 
description of the lowest mode of the 
spin dynamics was not rigorously derived from our multi-orbital Hamiltonian 
Eqs.~(\ref{hamkin}-\ref{hamint}). The acoustic mode reflects the frustrated nature of 
the magnetism within the block-OSMP phase. Also, the $J_1$--$J_2$ model may be
relevant in a wide range of interaction $U$ within the OSMP phase, beyond the block 
ordering region $0.4\lesssim U/W\lesssim 1.5$. For example, previous results showed 
that in the range $1.5\lesssim U/W\lesssim 20$ the system is in a 
ferromagnetic-OSMP~\cite{Rincon2014-1}, where the spins within the localized orbital 
$\gamma=2$ have ferromagnetic ordering. Clearly, a $J_1$--$J_2$ model with small or 
vanishing $J_2$ will also exhibit a similar ordering. Finally, note that although the 
alternative $S=1$ toy model is useful in the description of elementary states of the 
block-OSMP system, its validity is limited for the dynamical spin response: it is well 
known that the dynamical SSF of the $S=1$ AFM Heisenberg model exhibits ``sharp'' 
magnon lines, in contrast to the $S(q,\omega)$ of the $S=1/2$ model that contains a 
continuum of excitation (at least at low-$\omega$), in agreement with our results for 
the three-orbital Hamiltonian.

{\bf Optical mode.} 
Let us now turn to the high-energy optical mode of the dynamical SSF spectrum. The 
states contributing to this mode are also triplet excitations. In the $L=4$ Lanczos 
analysis we found that this high-energy mode arises from a state of the form 
\mbox{$|\mathrm{O}\rangle_{\gamma=2}\simeq1/2(|\downarrow\uparrow\uparrow\downarrow\rangle_{\gamma=2}+|\uparrow\downarrow\downarrow\uparrow\rangle_{\gamma=2})$}. 
It is evident that $|\mathrm{O}\rangle$ breaks the FM magnetic unit cells present in 
the GS. Note, again, that the discussed states do not have doubly occupied or empty 
sites, reflecting the Mott nature of orbital $\gamma=2$. 
It should be also pointed out that using a small $L=4$ system with OBC we have found another 
state which contributes to the optical mode, i.e., 
\mbox{$|\widetilde{\mathrm{O}}\rangle_{\gamma=2}=1/2(|\downarrow\uparrow\downarrow\uparrow\rangle_{\gamma=2}+|\uparrow\downarrow\uparrow\downarrow\rangle_{\gamma=2})$}. However, such a state is not 
present in the system with periodic boundary conditions. 

\begin{figure}[!ht]
\includegraphics[width=1.0\columnwidth]{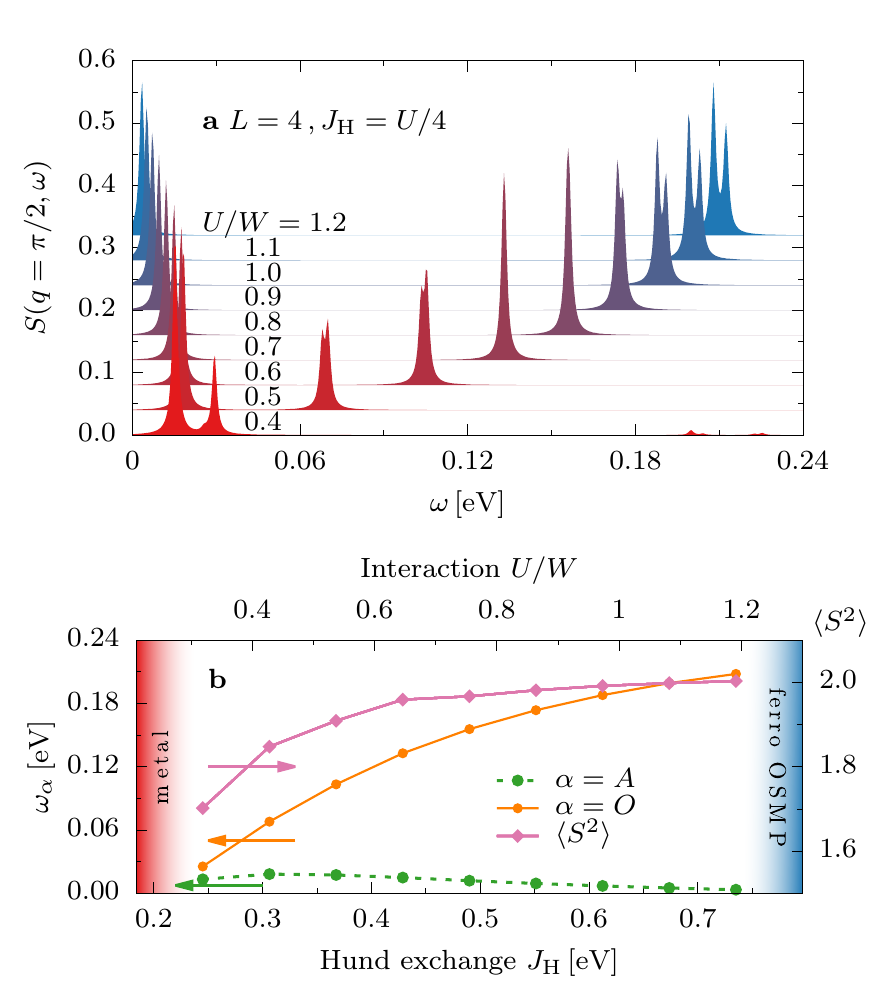}
\caption{{\bf Hund exchange dependence.} (a) Dynamical SSF at 
$q/\pi=1/2$ for various values of the interaction $U$, all within the block-OSMP 
phase, $U/W=0.4,\dots,1.2$ (bottom to top, with $0.04$ offset), at a fixed 
$J_{\mathrm{H}}/U=1/4$, calculated for $L=4$ using the Lanczos method. (b) Left $y$-axis: Frequency 
$\omega_\alpha=\epsilon_\alpha-\epsilon_{\mathrm{GS}}$ dependence of the acoustic and 
optical modes vs the value of the Hund exchange coupling $J_{\mathrm{H}}$. Right $y$-axis:
Magnetic moment $\langle S^2\rangle$ development within the block-OSMP phase.}
\label{fig:hund}
\end{figure}

To understand properly the optical mode it is not enough to focus solely on the 
localized orbital. A detailed analysis of the remaining ``metallic'' orbitals 
$\gamma=0,1$ indicates that: (i) the $|\mathrm{GS}\rangle$ and the 
$|\mathrm{A}\rangle$ states obey the Hund's rule: spins in different orbitals of the 
same site are ferromagnetically aligned [see Fig.~\ref{fig:conf}(a) for a schematic 
representation]. (ii) However, the $|\mathrm{O}\rangle$ states, 
Fig.~\ref{fig:conf}(c), do not fulfill this rule because part of the spins are 
antiferromagnetically aligned. As a consequence, the main difference in energy between 
the $|\mathrm{GS}\rangle$ and $|\mathrm{O}\rangle$ arises from the local (on-site) 
Hund exchange portion of the electronic interaction. We confirm this by calculating 
separately the expectation values of all terms contributing to the Hamiltonian (see 
Methods section). The main difference between the energy of the $|\mathrm{GS}\rangle$ 
and $|\mathrm{A}\rangle$ arises from the kinetic portion. On the other hand, the 
difference in $|\mathrm{O}\rangle$ originates, as expected, from the Hund coupling 
part of the interaction energy. The local on-site nature of the optical mode is also 
visible in the orbital resolved SSF. In Fig.~\ref{fig:omega}(c) we present the spin 
correlations between different orbitals at different sites, i.e. $S_{\mathrm{mix}}$. 
As clearly visible, the $S_{\mathrm{mix}}(q\to\pi)\to0$, indicating a drastic 
reduction of spectral weight at large momentum. These findings indicate that the 
optical mode is not present in the inter-orbital inter-site spin correlations. As a 
consequence, the only remaining possibility of the origin of the optical mode are the 
intra-site fluctuations between orbitals. Our investigation of orbital resolved SSF [
see Fig.~\ref{fig:omega}(c)] shows that each orbital contributes to the optical mode 
with a similar weight. Finally, the lack of momentum dependence of the optical mode (
at least for $q/\pi>1/2$) suggests that such excitations are local (on-site) 
fluctuations of spin between different orbitals at the same site.

\begin{figure*}[!ht]
\includegraphics[width=1.0\textwidth]{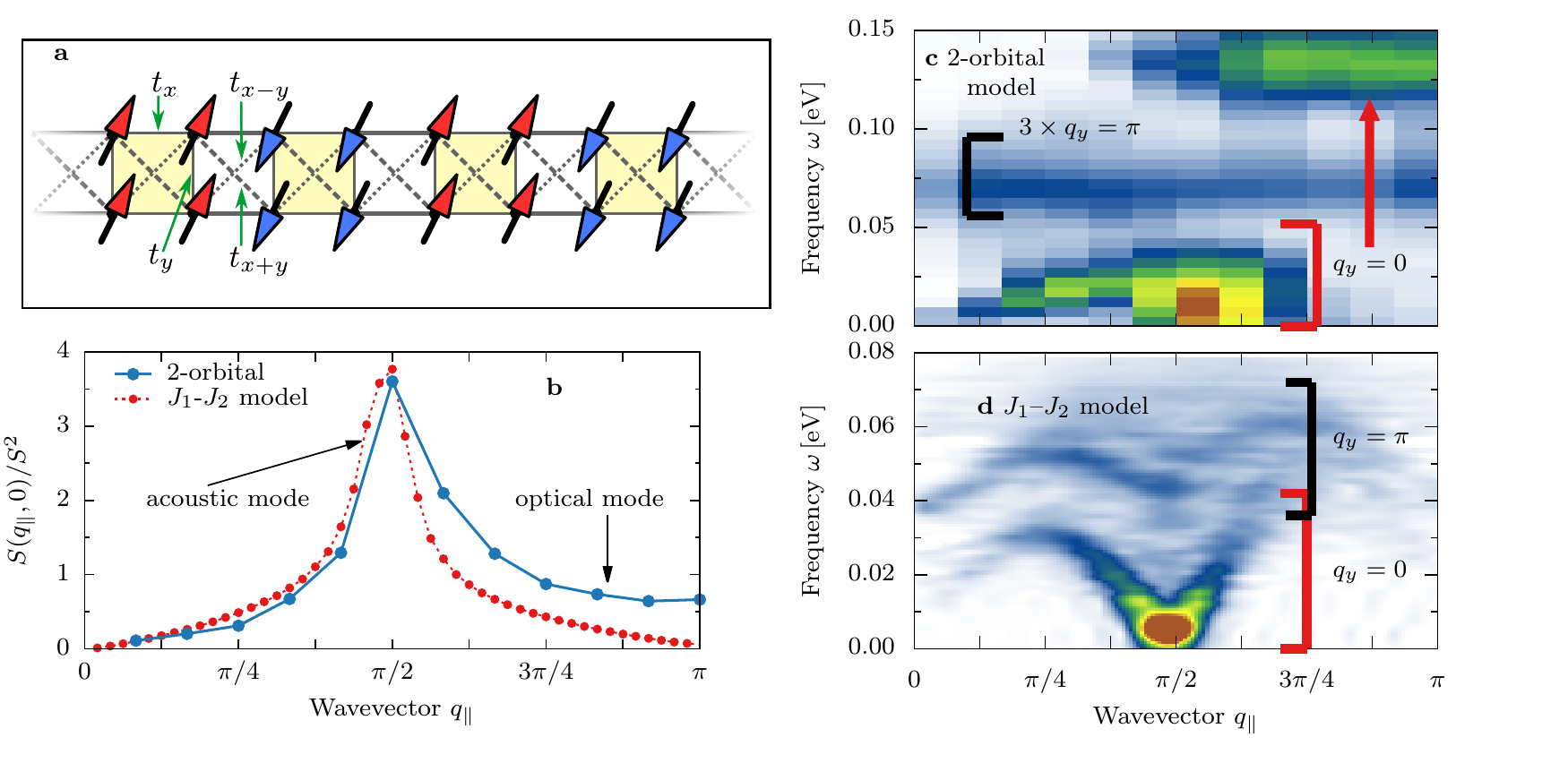}
\caption{{\bf Spin dynamics within the ladder geometry.} (a) Schematic representation 
of the two-orbital two-leg ladder system used in our analysis (see Methods section for 
details). Arrows depict the 2$\times$2 block state. (b) Static SSF in the $q_y=0$ 
sector for both bthe $J_1$-$J_2$ and multi-orbital models on a ladder geometry. (c,d) 
Dynamical SSF of the (c) two-orbital and (d) $J_1$-$J_2$ spin models on a two-leg 
ladder geometry (using $L=12$ and $L=48$ rungs, respectively).}
\label{fig:ladd}
\end{figure*}

In addition, we have shown that the frequency 
$\omega_{\mathrm{O}}=\epsilon_{\mathrm{O}}-\epsilon_{\mathrm{GS}}$ of the 
corresponding $|\mathrm{O}\rangle$ excitation is directly proportional to the value of 
the Hund exchange $J_{\mathrm{H}}$, contrary to the $|\mathrm{A}\rangle$ excitation with energy 
$\omega_{\mathrm{A}}=\epsilon_{\mathrm{A}}-\epsilon_{\mathrm{GS}}$. In 
Fig.~\ref{fig:hund}(a) we present the dynamical SSF at $q/\pi=1/2$ for various values 
of $U$ within the block-OSMP calculated via the Lanczos method on  $L=4$ sites, at a 
fixed $J_{\mathrm{H}}/U=1/4$. Our results in Fig.~\ref{fig:hund}(b) indicate that this behavior 
is valid throughout the entire block-OSMP phase, $0.4\lesssim U/W \lesssim 1.5$.

{\bf Ladder geometry.} 
Finally, let us comment on the lattice geometry dependence of our results. In 
Fig.~\ref{fig:ladd} we present the SSF for the two-leg ladder two-orbital Hamiltonian 
introduced in Ref.~\cite{luo2013} for the BaFe$_2$Se$_3$ compound. The lattice is 
sketched in Fig.~\ref{fig:ladd}(a) and hopping values are given in the Methods 
section. It was previously shown~\cite{luo2013} that at density $\overline{n}=1.75/2$, 
$J_{\mathrm{H}}=U/4$, and $U/W_L=2$ with $W_L=3.82\,\mathrm{eV}$ the system is in an 
enlarged block phase, similar to the $2\times2$ block state of 
BaFe$_2$Se$_3$~\cite{Mourigal2015}. Before addressing specific results, it is 
important to remark that the DMRG numerical studies of multi-orbital ladders require 
expensive computations. This is because the inter-site inter-orbital hoppings behave 
effectively as long-distance hoppings in the equivalent one-dimensional 
representation, leading to larger entanglement for the ground state 
(see Supplementary Note~1 for details). The calculation of dynamical quantities is 
certainly a challenge and even the static expectation values have to be carefully 
analyzed with regards to the number of states kept (here $M=1000$ states are used). As 
a consequence, the results presented for the two-orbital two-leg ladder below may not 
be as accurate as those for the chains.

On a ladder, there are two separate contributions to the SSF arising from the bonding 
($q_y=0$) and antibonding ($q_y=\pi$) sectors. For the two-orbital two-leg ladder 
results, presented in Fig.~\ref{fig:ladd}(b), we find a $q_y=0$ dispersive mode at low-
$\omega$, with a continuum of spin excitations similar to the acoustic mode of the 
chain geometry. At $\omega\simeq0.075\,\mathrm{eV}$ we find an energy narrow $q_y=\pi$ 
mode. According to our analysis of the 1D system, a similar spectrum can be found in 
the $J_1$-$J_2$ model on the ladder with FM rung coupling $J_{\perp}=J_1$, see 
Fig.~\ref{fig:ladd}(c-d). Both the $J_1$-$J_2$ spin model and multi-orbital model on the 
ladder studied here exhibit the 2$\times$2 block state, i.e. AFM coupled blocks of 
four FM aligned spins on two neighbouring rungs [see Fig.~\ref{fig:ladd}(a)]. Such a 
state has a peak in the static SSF at $q_x=\pi/2$ in the bonding contribution 
($q_y=0$), see Fig.~\ref{fig:ladd}(c). Note that the maximum of the acoustic mode 
appear at $\omega\ne0$, which suggests a non-zero spin gap, common in ladders. 
Finally, at higher frequencies ($\omega\simeq0.13\,\mathrm{eV}$) in the $q_y=0$ sector 
we find a flat mode of excitations, similar to the optical mode present in the chain 
analysis. It is again evident that the latter is not captured by the $J_1$-$J_2$ model.

\section{Discussion}
Let us compare the INS data for BaFe$_2$Se$_3$ reported in 
Ref.~\onlinecite{Mourigal2015} against our results. Note that this compound is 
insulating~\cite{Caron2012}, while our system Eqs.~(\ref{hamkin}-\ref{hamint}) for the 
parameters considered in this work, $U/W=0.8$ and $J_{\mathrm{H}}/U=1/4$, is a (bad) metal in the 
block-OSMP phase, becoming insulator only for $U/W\gtrsim 1.5$ in the 
ferromagnetic-OSMP phase~\cite{Li2016}. 
Our Hamiltonian reproduces the OSMP state and the magnetic block phase of BaFe$_2$Se$_3$,
and although the charge dynamics of our model does not capture the experimentally observed 
insulating nature of the real material, it is still appealing to study the spin physics.
The lack of other multi-orbital models that can reproduce both the spin and charge sector
of low-dimensional iron selenides makes it appealing to carry out detailed theoretical calculations 
of the spin dynamics within this model and compare with the experiments.

Within the spin-wave theory the low-$\omega$ portion of the INS spectra was 
interpreted~\cite{Mourigal2015} as a dispersive mode which reflects the frustrated 
nature of the $\pi/2$-order. In addition, the high-energy optical modes were interpreted 
as local excitation of spins within the 2$\times$2 plaquette. A similar rationale was 
used to explain the INS result of the doped compound 
Rb$_{0.89}$Fe$_{1.58}$Se$_2$~\cite{wang2011}. The spin-wave theory of 
BaFe$_2$Se$_3$ reproduces~\cite{Mourigal2015} all of the modes and also properly 
captures the frequency bandwidth of the spin excitations. However, only $\sim2/3$ of 
the total spectral weight expected for localized 3$d$ electrons is obtained. Also note 
that within the considered spin models of Ref.~\onlinecite{Mourigal2015,wang2011} 
unphysically large dimerization spin-exchange couplings 
are required~\cite{Bao2014,Popovic2015} to stabilize the $\pi/2$ spin pattern.

From the perspective of our results, the interpretation of the INS spin spectra of 
low-dimensional ladder iron chalcogenides is different from spin-wave theory. The 
latter assumes that all excitations occur between localized spins, while in our system 
we have a mixture of localized and itinerant electrons. Moreover, as shown above, the 
SSF of multi-orbital systems not only contains dispersive acoustic modes but also 
local excitations controlled by the Hund exchange, at least within the block-OSMP. The 
inter-orbital nature of such modes cannot be properly captured by localized Heisenberg 
models. Our results, on both chain and ladder geometries, indicate that spin models 
can only properly capture dispersive modes resulting from the peculiar spin order of a 
given phase as in the $\pi/2$ state of BaFe$_2$Se$_3$. However, we argue that only one 
of the low lying optical modes of this compound arises from a weakly dispersive (
probably beyond experimental resolution of powder sample) $q_y=\pi$ excitation. Within 
our interpretation of the SSF spectra, the second optical mode is of a different 
nature, involving inter-orbital spin fluctuations on each site. Such a picture is 
consistent with our multi-orbital ladder results.

Concerning the spectral weight, for the chosen parameters $U/W=0.8$ and
$J_{\mathrm{H}}/U=1/4$ in Eqs.(\ref{hamkin}-\ref{hamint}) we observe the magnetic moment 
$\langle S^2\rangle \sim 2$ (maximal possible for $\overline{n}=4/3$). This is 
consistent with previous Hartree-Fock calculations \cite{luo2013} of the 
block-OSMP phase within a five-orbital ladder system, which reported 
$\langle S^2\rangle \sim 6$ for $\overline{n}=6/5$ (again the maximal value). 
As a consequence, our results do not reproduce the missing spectral weight observed in
experiments~\cite{Mourigal2015}. However, the magnetic moments evolve 
within the block-OSMP~\cite{Rincon2014-1} 
(see also Supplementary Note~2 for additional results) and only saturate to its maximal 
value at $U/W\gtrsim0.6$, namely in the middle of the block-phase. Since the 
exact value of $U$ and $J_{\mathrm{H}}$ are not know for BaFe$_2$Se$_3$, it is possible in 
theoretical investigations to stabilize the block-OSMP phase with a reduced 
$\langle S^2\rangle<2$ [see Fig.~\ref{fig:hund}(b)].
Moreover, note that recently it was argued \cite{mannella} that insufficient 
energy (time) resolution in
INS experiments produces moments that can be smaller than the actual instantaneous moments.
In this context, faster x-ray based techniques such as photoemission spectroscopy (PES), 
x-ray absorption spectroscopy (XAS), and x-ray emission spectroscopy (XES) 
are needed to resolve this issue.

In conclusion, we have investigated the dynamical spin structure factor of a 
one-dimensional three-orbital Hubbard model in the block orbital selective Mott phase, 
as well as a ladder two-orbital Hubbard model also in a similar block state. This has 
been a computationally demanding effort even with the powerful DMRG, and to our 
knowledge this is the first time that results of this quality are produced. We have 
shown that our Hamiltonian captures nontrivial features of a broad family of 
low-dimensional iron chalcogenides, in particular for the ladder BaFe$_2$Se$_3$ 
compound for which $\pi/2$--block order was reported. We have found two different 
types of modes in the spin spectra: (i) low-frequency dispersive (acoustic) spin 
excitations and (ii) optical dispersionless excitations at higher energy. The acoustic 
band reflects the nature of magnetic order of the system, namely for the block-OSMP 
the frustrated $\pi/2$-ordering can be captured by the quantum $J_1$-$J_2$ frustrated 
Heisenberg model, as also shown here. The optical band arises from on-site 
inter-orbital spin fluctuations controlled by the Hund exchange coupling. Finally, our 
1D dynamical SSF is in qualitative agreement with the powder INS spectrum of 
BaFe$_2$Se$_3$ (see Supplementary Note~3). Although the latter has only a quasi-1D geometry, with 
small but nonzero couplings perpendicular to the ladder, the $\omega$ dependent 
spectra should be dominated by the predominantly 1D nature of the system. As a 
consequence, the location in momentum and energy space is properly resolved by our 
model Hamiltonian Eqs.~(\ref{hamkin}-\ref{hamint}) for both of the modes.

Our results are general and should apply to a variety of block states in multi-orbital 
quasi-1D systems. They should all contain an acoustic band (with pitch wavevector 
compatible with the size of the magnetic block), a strong asymmetry in the 
distribution of weight of this acoustic band in different portions of the Brillouin 
zone, and optical modes with at least one of them related to atomic transitions 
regulated by the Hund coupling. 

\section{Methods}

\noindent {\bf DMRG method.}
The Hamiltonians discussed here were studied using primarily the density matrix 
renormalization group (DMRG) method~\cite{white1992,schollwock2005} within the 
single-center site approach~\cite{white2005}, where the dynamical correlation 
functions are evaluated via the 
dynamical DMRG~\cite{jeckelmann2002,benthein2007,nocera2016} i.e. calculating spectral 
functions directly in frequency space with the correction-vector 
method~\cite{kuhner1999} with Krylov decomposition~\cite{nocera2016}. The computer 
package \textsc{DMRG++} developed at ORNL was used. For a chain geometry, in both 
stages of the DMRG algorithm, we keep up to $M=800$ states. This allow us to simulate 
accurately system sizes up to $L=24$ sites for dynamical quantities (truncation 
$<10^{-8}$ for all frequencies $\omega$) and $L=32$ for static quantities (truncation 
$<10^{-10}$ for the GS). For the ladder geometry results, we use a standard two-site 
central block approach with $M=1000$ states (truncation $<10^{-3}$, showing that the 
two-leg ladder two-orbital results are qualitatively correct, because of its close resemble 
to the rest, but their quantitative accuracy can be further improved in future efforts). 
In the Supplementary Note~1 we present the scaling of our results with system size 
$L$, number of states kept $M$, and broadening $\eta$ of Eq.~(\ref{stfa}).

\noindent {\bf Dynamical SSF.} 
The zero temperature, $T=0$, total spin structure factor (SSF) $S(q,\omega)$ is 
defined as:
\begin{eqnarray}
S(q,\omega)&=&\frac{1}{\pi}\sqrt{\frac{2}{L+1}}\sum_{\ell=1}^{L}
\sin(q\ell)\sin(qL/2)\times\nonumber\\
&&\mathrm{Im}\,\langle\mathrm{GS}|\tilde{S}_{\ell}\,
\frac{1}{\omega^{-}-(H-\epsilon_{\mathrm{GS}})}\,
\tilde{S}_{L/2}|\mathrm{GS}\rangle\,,
\label{stfa}
\end{eqnarray}
with $\omega^{-}=\omega-i\eta$, and $|\mathrm{GS}\rangle$ is the ground state with 
energy $\epsilon_{\mathrm{GS}}$. In the above equation 
$\tilde{S}_{\ell}=\sum_{\gamma}S_{\ell,\gamma}$ is the total spin on site $\ell$ for 
the total SSF $S(q,\omega)$, or $\tilde{S}_{\ell}=S_{\ell,\gamma}$ for the 
orbital resolved SSF $S_{\gamma\gamma^\prime}(q,\omega)$.

Furthermore, in the above equation we adopted the wave-vector definition appropriate 
for open boundary conditions (OBC), i.e. $q=k\pi/(L+1)$ with $k=1,\dots,L$. As a 
consequence, in this work we used approximate (exact in the thermodynamic limit 
$L\to\infty$) values of the wave-vectors, e.g., $q=\pi\equiv\pi L/(L+1)$. 

\noindent {\bf Localized basis representation.}
The eigenstates $|\phi\rangle$ of the three orbital system can be written as
\begin{eqnarray}
|\phi\rangle &=& \sum_{n=1}^{64^L} c_n | n\rangle\nonumber\\
&=&\sum_{n_0=1}^{4^L} \sum_{n_1=1}^{4^L} \sum_{n_2=1}^{4^L} c(n_0,n_1,n_2)\,
|n_0\rangle\otimes|n_1\rangle\otimes|n_2\rangle\,,
\end{eqnarray}
where $|n\rangle$ represent the orthonormal basis (particle configuration) of all 
orbitals and $|n_\gamma\rangle$ (with $\gamma=0,1,2$) represents the particle 
configuration on given orbital $\gamma$. Note that 
$\sum_n c^2_n=\sum_{n_1,n_2,n_3}c^2(n_1,n_2,n_3)=1$ and 
$\langle n_\gamma|n^\prime_{\gamma^\prime}\rangle=\delta_{nn^\prime}\delta_{\gamma\gamma^\prime}$. 
One can rewrite the above equation as
\begin{equation}
|\phi\rangle = \sum_{j=1}^{4^L} |\tilde{c}_j\rangle\otimes|j\rangle_{\gamma=2}\,,
\end{equation}
where $j\equiv n_2$ represents - within OSMP - the localized orbital and
\begin{equation}
|\tilde{c}_j\rangle=\sum_{n_0=1}^{4^L} \sum_{n_1=1}^{4^L} c(n_0,n_1,n_2)\,
|n_{0}\rangle\otimes|n_{1}\rangle
\end{equation}
are vectors. The set of $\{|\tilde{c}_j\rangle\}$ vectors represent an orthogonal 
vector-space with $\sum_{j} \langle \tilde{c}_j|\tilde{c}_j\rangle=1$. Finally, the 
weight of the $|j\rangle_{\gamma=2}$ configuration in the $|\phi\rangle$ eigenstate is 
given by the norm of the $|\tilde{c}_j\rangle$ vector, i.e., 
$\langle \tilde{c}_j|\tilde{c}_j\rangle=||\tilde{c}_j||\equiv\tilde{c}_j^2$.

\noindent {\bf Energy contribution.} 
In Table~\ref{entable} we present the expectation values of the several terms present 
in the Hamiltonian Eqs.~(\ref{hamkin}-\ref{hamint}) for the ground state and also 
states which contribute to the acoustic and optical modes.

\begin{table}[!ht]
\centering
\caption{{\bf Energy contributions.} Kinetic, intra- and inter-orbital interaction, 
Hund, and pair-hopping energy contributions to the energy of given eigenstates. The 
last column shows the difference between $|\mathrm{GS}\rangle$ and states within the 
acoustic (red color) and optical (green color) modes. Results are obtained for $L=4$ 
and $U/W=0.8$, using the Lanczos method. All numbers in units of eV.}
\label{entable}
\begin{tabular}{c|c|c|c|c|c|c|c|}
\cline{2-8}
                                            & $\epsilon_{\mathrm{k}}$    & $\epsilon_{\mathrm{U}}$ & $\epsilon_{\mathrm{U}^\prime}$ & $\epsilon_{\mathrm{H}}$ & $\epsilon_{\mathrm{P}}$ & Total & $\omega_\alpha$ \\ \hline
\multicolumn{1}{|c|}{$|\mathrm{GS}\rangle$} & $-0.027$                   & $8.006$                 & $15.280$                       & $-1.055$ & $-0.010$     & $22.194$                &                         \\ \hline
\multicolumn{1}{|c|}{$|\mathrm{A}\rangle$}  & $0.007$                    & $7.993$                 & $15.280$                       & $-1.065$                & $-0.009$                & $22.206$ & $0.012$      \\ \hline
\multicolumn{1}{|c|}{$|\mathrm{O}\rangle$}  & $-0.031$                   & $8.081$                 & $15.262$                       & $-0.946$ & $-0.016$     & $22.350$                & $0.156$ \\ \hline
\end{tabular}
\end{table}

\noindent {\bf Two-orbital two-leg ladder Hamiltonian.} The symmetric hoppings for the 
two-orbital two-leg ladder system are defined~\cite{luo2013} in orbital space as 
follows [see sketch in Fig.~\ref{fig:ladd}(a)]:

\begin{equation*}
t_x=
\begin{pmatrix}
0.14769 & 0 \\
0 & 0.27328
\end{pmatrix}\,,
\qquad
t_y=
\begin{pmatrix}
0.28805 & 0.01152 \\
0.01152 & 0.00581
\end{pmatrix}\,,
\end{equation*}
\begin{equation*}
t_{x\pm y}=
\begin{pmatrix}
-0.21166 & \mp0.08430 \\
\mp0.08430 & -0.18230
\end{pmatrix}\,,
\end{equation*}
all expressed in units of $\mathrm{eV}$. The interaction portion of the Hamiltonian is 
the same as in the 1D system Eq.~(\ref{hamint}).

\noindent {\bf Data availability.} The data that support the findings of this study are available from the corresponding author upon request.

\noindent {\bf Code availability.} Computer codes used in this study are available 
at \url{https://g1257.github.io/dmrgPlusPlus/}.


\section{Acknowledgments}
J.H, A.M., and E.D. were supported by the US Department of Energy (DOE), Office of Science, Basic Energy Sciences (BES), Materials Sciences and Engineering Division. N.K. was supported by the National Science Foundation Grant No. DMR-1404375. The work of G.A. was conducted at the Center for Nanophase Materials Science, sponsored by the Scientific User Facilities Division, BES, DOE, under contract with UT-Battelle.

\section{Author contribution}
J.H. and E.D. planned the project. J.H. performed all DMRG calculations for the 
multi-orbital Hubbard model, N.K. performed all Lanczos and $J_1$-$J_2$ model calculations, 
while A.N. and G.A. developed the \textsc{DMRG++} computer program. J.H., A.M., and E.D. 
wrote the manuscript. All co-authors provided comments on the paper.

\section{Additional information}
\noindent {\bf Supplementary Information} accompanies this paper.

\noindent {\bf Competing Interests} The authors declare no competing interests.

\clearpage
\appendix
\setcounter{figure}{0}
\setcounter{equation}{0}
\newcommand{\rom}[1]{\uppercase\expandafter{\romannumeral #1\relax}}
\renewcommand{\citenumfont}[1]{S#1}
\renewcommand{\bibnumfmt}[1]{[S#1]}
\renewcommand{\thefigure}{S\arabic{figure}}
\renewcommand{\theequation}{S\arabic{equation}}
\begin{widetext}
\begin{center}
{\bf \uppercase{Supplementary Information}} for:\\
\vspace{5pt}
{\bf \large Spin dynamics of the block orbital-selective Mott phase}\\
\vspace{5pt}
by J. Herbrych, {\it et al.}
\end{center}

\section{Supplementary Note 1. Numerical details}

In Supplementary Fig.~\ref{fig:dmrg} we present the parameter dependence of our dynamical DMRG 
calculations for a fixed frequency $\omega=0.03\,\mathrm{[eV]}$ (namely, ``inside'' the 
acoustic mode) and $L=16$ sites ($48$ orbitals). In panel (a) we present the broadening 
$\eta$ dependence of our calculations [Eq.~(4) of the main text]. It is clear from the 
figure that all features are properly resolved for the considered 
$\eta/\delta\omega=2$. In Supplementary Fig.~\ref{fig:dmrg}(b) we present the number of states kept 
$M$ dependence of our findings. We conclude that at a fixed $\eta$ and $L$, the results 
do not change appreciably for $M\gtrsim 800$.

\begin{figure}[!ht]
\includegraphics[width=0.5\textwidth]{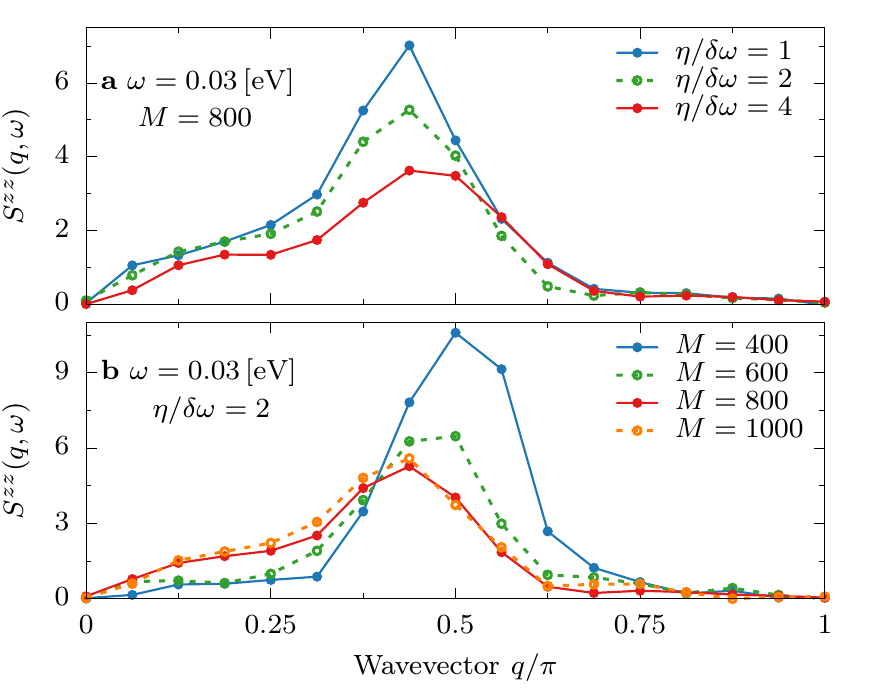}
\caption{{\bf Parameter dependence of dynamical-DMRG simulations.} (a) Broadening 
$\eta$ and (b) number of states kept $M$ dependence corresponding to 
$\omega=0.03\,\mathrm{[eV]}$ and $L=16$ sites. In all simulations of the main text we 
use $\eta/\delta\omega=2$ and $M=800$.}
\label{fig:dmrg}
\end{figure}

\begin{figure}[!ht]
\vspace{10pt}
\includegraphics[width=0.5\textwidth]{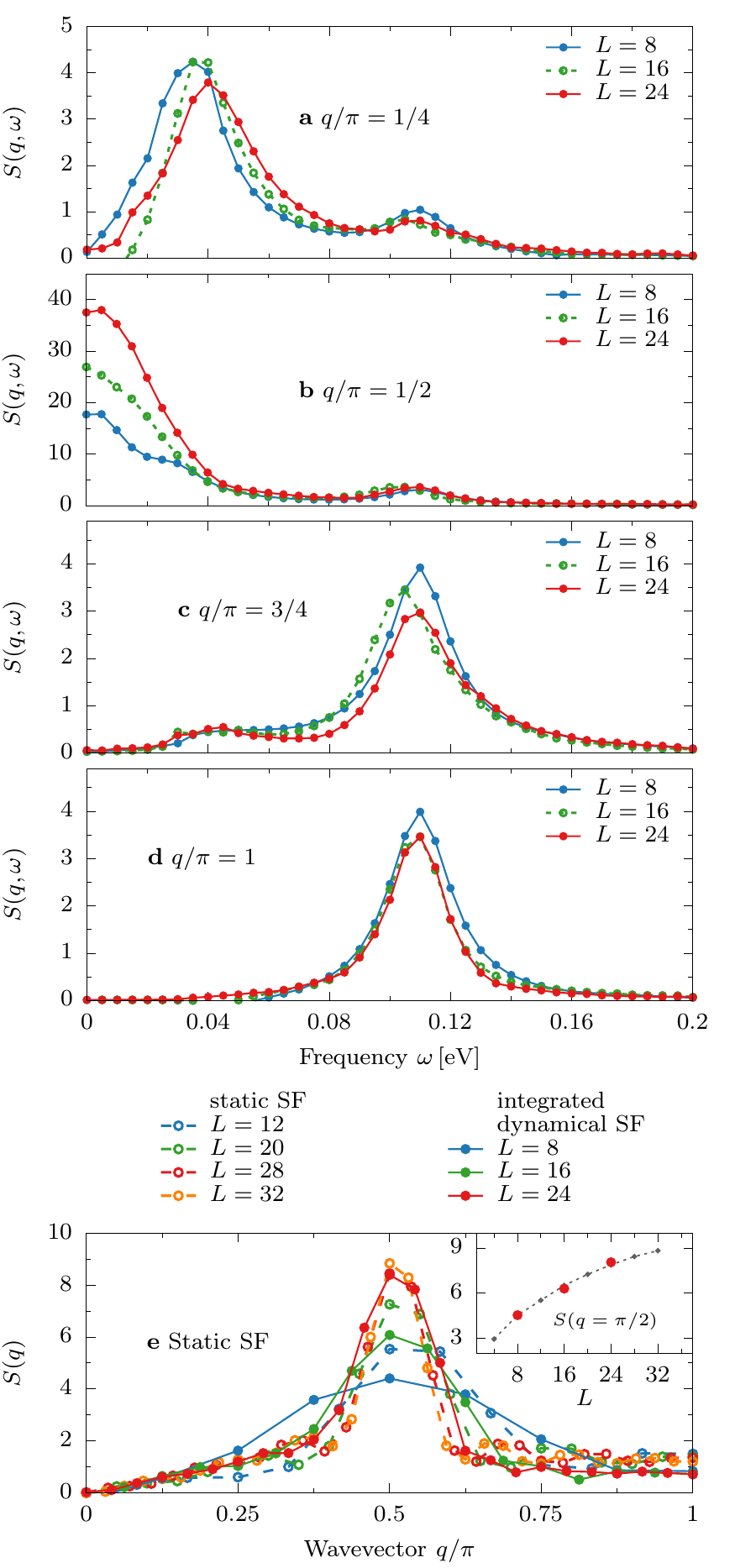}
\caption{{\bf Finite-size analysis.} (a-d) Size $L$ dependence of the 
frequency-resolved dynamical SSF for $q/\pi=1/4,1/2,3/4,1$, as calculated with 
$\eta/\delta\omega=2$ and $M=800$. (e) $L$-dependence of the static SSF. Open points 
represent the results obtained as the expectation value of the GS, while solid points 
are obtained from the integral over the frequency (see main text for details). Inset 
illustrates the quasi-long-range nature of block $\pi/2$ ordering, with a signal 
intensity growing with $L$.}
\label{fig:size}
\end{figure}

In Supplementary Fig.~\ref{fig:size}(a-d) we present the finite-size analysis at several momenta $q$ 
cuts through the dynamical SSF. At large $q/\pi\geq 3/4$, the results do not depend on the 
system size $L$ because for this momentum only the optical mode is present in the 
spectrum. Since the excitations within this mode are local, the system size (and also 
dimensionality of the lattice) should not play a crucial role. On the other hand, at 
$q\leq \pi/2$ the results depend more on the system size with maximal variation at 
$q/\pi=1/2$. However, this dependence does not change the main findings of our work and 
it merely reflects the quasi-long-range nature of the block 
ordering~\cite{SRincon2014-1}. This can be understood simply from the $L$-scaling of 
the static $S(q=\pi/2)$ shown in the inset of Supplementary Fig.~\ref{fig:size}(e). For completeness in 
Supplementary Fig.~\ref{fig:size}(e) we show the $L$ dependence of the full momentum $q$ resolved 
static SSF.

Let us finally comment on the accuracy of our results for the multi-orbital ladder 
geometry. Different from the chain setup, where the three orbitals where treated as a 
single site with a local Hilbert space of $64$ states, the ladder results were obtained 
using a $12\times2\times2$ (rungs $\times$ legs $\times$ orbitals) lattice with a local 
Hilbert space of $4$ states. Although such a setup have smaller memory requirements, 
the entanglement area law~\cite{Avella2013} heavily influences the accuracy of our 
results. The latter is a consequence of a large number of long-range connections (up to 
$7$ nearest-neighbours). In Supplementary Fig.~\ref{fig:dmrg_ladder}, we present the system size $L$ 
and states $M$ scaling of the results presented in Fig.~7 of the main text. In panel 
(a) we present the finite-size analysis of the static SSF in the bonding sector, 
$q_y/\pi=0$, for the $M=1000$ states kept. The system size analysis of the ladder 
results is consistent with the findings for chains, namely the acoustic mode has size 
dependence, while the optical mode does not. In summary, while we are confident that 
our results for ladders capture the essence of the problem, including the existence of 
acoustic and optical bands and quite different weights in different portions of the 
Brillouin zone, only further (very demanding) work can achieve the same accuracy as 
shown here for chains.

\begin{figure}[!ht]
\includegraphics[width=0.5\textwidth]{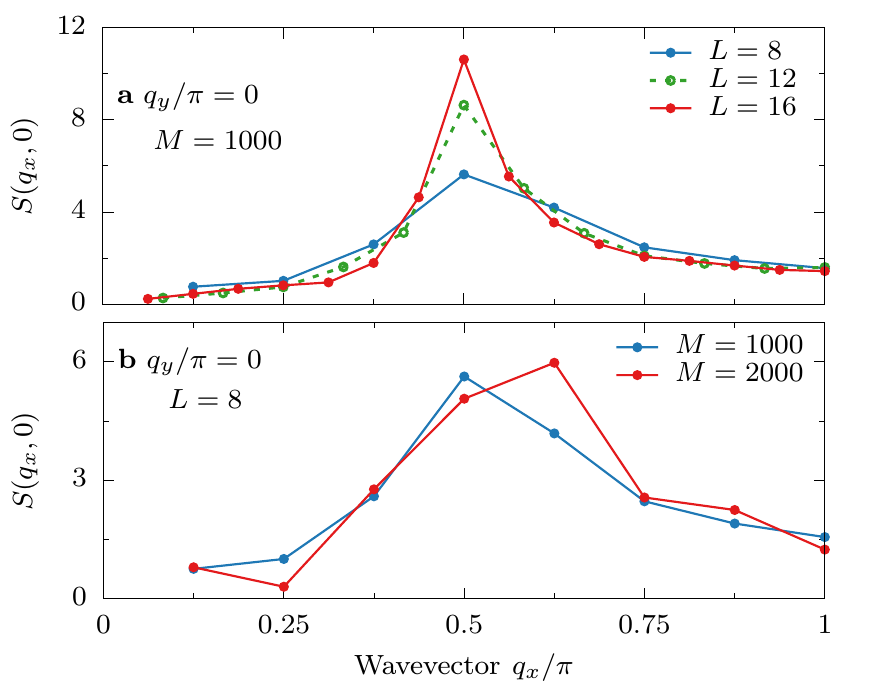}
\caption{{\bf Ladder geometry analysis.} (a) Finite size $L$ and (b) number of states 
kept $M$ scaling of the static SSF in the bonding sector, $q_y/\pi=0$.}
\label{fig:dmrg_ladder}
\end{figure}

\section{Supplementary Note 2. Magnetic moment evolution.}

In Supplementary Fig.~\ref{fig:s2} we present the evolution of the local magnetic moment
 $\langle S^2\rangle$ 
within the block-orbital selective Mott phase. This local moment can be obtained 
from the sum-rules of spin-spin correlation functions, i.e.,
\begin{equation}
S(q)=\frac{1}{\pi}\int\mathrm{d}\omega\,S(q,\omega)\,,\quad
\langle S^2\rangle=\frac{1}{L} \int\mathrm{d}q\,S(q)\,.
\end{equation}
Note that the above equations allow to relate the total spectral weight of INS data
with the value of the local spin via $\langle S^2\rangle=S(S+1)$. The results presented in
Supplementary Fig.~\ref{fig:s2} are obtained from the integration of the static structure factor $S(q)$.
As clearly visible, the magnetic moments start to develop already in the paramagnetic 
(metallic) phase \cite{SRincon2014-1} and are stabilized to its maximal value 
$\langle S^2\rangle$ ($S=1$ for $\overline{n}=4/3$) in the middle of the block-OSMP.

\begin{figure}[!ht]
\includegraphics[width=0.5\textwidth]{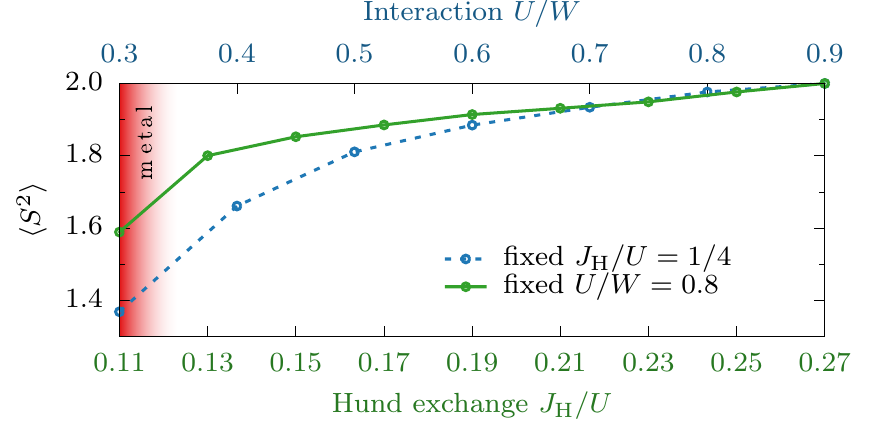}
\caption{{\bf Magnetic moment.} Evolution of the local magnetic moment $\langle S^2\rangle$ 
within the block-OSMP. The solid line (lower $x$-axis) represents results for fixed 
value of interaction $U/W=0.8$ and various value of $J_{\mathrm{H}}/U$. The dashed line (upper 
$x$-axis) represents results at fixed $J_{\mathrm{H}}/U=1/4$ and for various values of $U/W$.
The results were obtained using a DMRG method with parameters $L=16$ ($48$ orbitals), $M=800$.}
\label{fig:s2}
\end{figure}

\section{Supplementary Note 3. Comparison of DMRG results with powder experiment.}

\begin{figure}[!ht]
\includegraphics[width=0.5\textwidth]{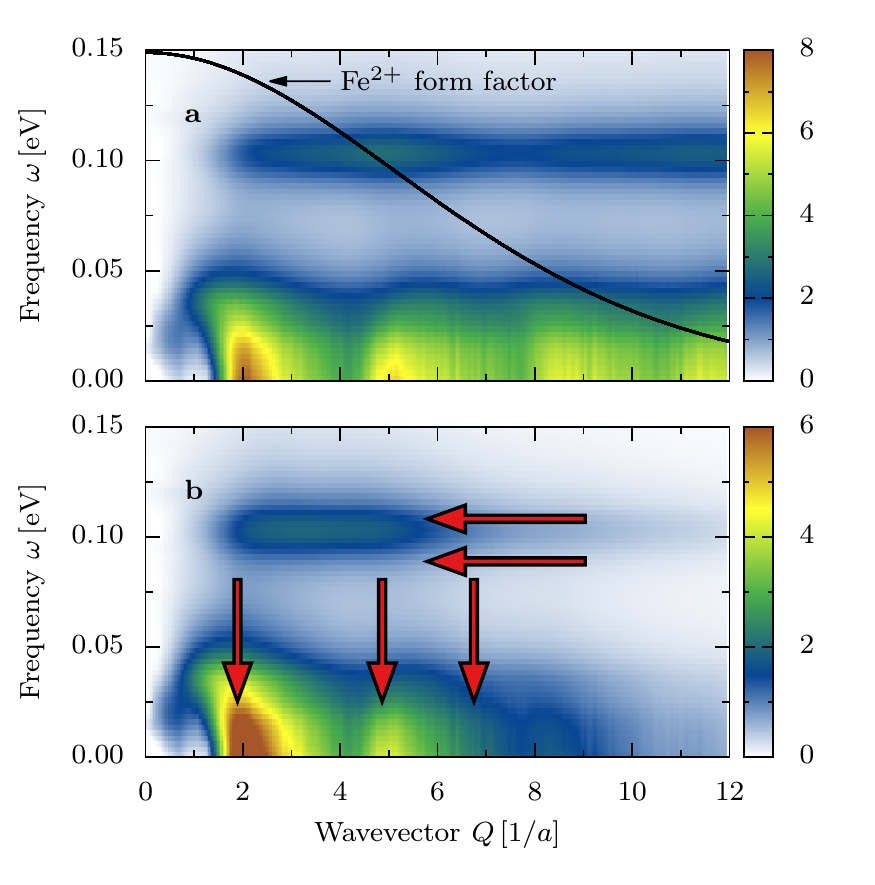}
\caption{{\bf Powder spectrum.} (a) Spherical average of the dynamical SSF. The black 
solid line represents the magnetic form factor $F(Q)^2$ of the Fe$^{2+}$ 
ions~\cite{Sbrown}. (b) Spherical average of the dynamical SSF convoluted with the 
form factor $F(q)$ relevant for a direct comparison with the BaFe$_2$Se$_3$ INS 
results~\cite{SMourigal2015}. Red arrows indicate the position of maximum intensities 
in the INS spectrum. See text for details.}
\label{fig:U08powd}
\end{figure}

Although BaFe$_2$Se$_3$ is a quasi-1D compound, the finite $\omega$-dependent 
properties should be dominated by the 1D nature of the ladder lattice (while, e.g., 
d.c. transport is more subtle). It is therefore appropriate to directly compare our 
dynamical SSF to experimental findings. Since the latter is obtained using a powder 
sample, our results presented in Fig.~2 of the main text have to be averaged over all 
spherical angles~\cite{Stomiyasu2009}. Furthermore, to qualitatively compare to the 
inelastic neutron scattering (INS) data we must incorporate in the analysis the 
momentum dependent magnetic form factors $F(Q)$ of the spin carriers, namely the 
Fe$^{2+}$ ions. Here we assume a gyromagnetic ratio $g=2$ (spin-only scattering). The 
functional form of the former can be taken from crystallography tables~\cite{Sbrown}. 
In Supplementary Fig.~\ref{fig:U08powd}(a) we present the powder average of our spectra. Several 
interesting general features can be inferred: (i) using realistic 
values~\cite{SMourigal2015} for the Fe-Fe distance such as $2.7$~\AA, remarkably we 
obtain a nearly perfect agreement for the position of the acoustic mode. The leading 
INS signal is centered at $Q\simeq 0.7$~($1/$\AA), followed by peaks at $1.8$~($1/$\AA) to 
$2.5$~($1/$\AA) with smaller intensity [indicated by vertical red arrows in Supplementary 
Fig.~\ref{fig:U08powd}(b)]. (ii) 
The neutron spectrum gives three flat (momentum-independent) bands of spin exactions: 
two of them are centered approximately at $\omega\sim0.1$~eV ($\omega_1=0.0889$~eV and 
$\omega_2=0.1082$~eV, depicted as horizontal red arrows in 
Supplementary Fig.~\ref{fig:U08powd}(b)), 
while the third one is positioned at $\omega_3=0.198$~eV. Our 1D results yield only 
one optical mode centered at $\omega\simeq0.105$~eV in accord with the most pronounced 
peak within the INS spectrum. This qualitative agreement indicates that our model is 
able to capture the nontrivial nature of the frustrated magnetism of BaFe$_2$Se$_3$, 
and that the studied parameter range of our Hamiltonian is valid for the whole 123 
family.


\end{widetext}
\end{document}